\documentclass[sigplan,10pt]{acmart}
\renewcommand\footnotetextcopyrightpermission[1]{}
\AtBeginDocument{%
  }

\usepackage[ruled,vlined,linesnumbered]{algorithm2e}
\usepackage{graphicx}
\usepackage{subcaption}
\usepackage[normalem]{ulem}
\usepackage{comment}
\usepackage{multirow}
\usepackage{amsthm}
\usepackage{enumitem}
\usepackage{booktabs}
\usepackage{tabularx}

\theoremstyle{definition}

\newcommand{\para}[1]{\vspace{0.5em}\noindent {\bf #1}}
\newcommand{\company}{Provider A}
\newcommand{\tool}{Astragalus}
\newcommand{\acr}{ACR}
\newcommand{\sbfl}{SBFL}
\newcommand{\fix}{update}
\ifdefined\revisionhighlight
  \long\def\revision#1{{\color{blue} #1}}
\else
  \long\def\revision#1{#1}
\fi

\ifdefined\commentenabled
  \long\def\comments#1{{\color{blue} [#1]}}
  \long\def\todo#1{{\color{blue} [Todo: #1]}}
  \long\def\zp#1{{\color{red} [Todo: #1]}}  
  \long\def\jiawei#1{{\color{blue} [Jiawei: #1]}}
  \long\def\lx#1{{\color{orange} [Xu: #1]}}
  \long\def\gu#1{{\color{olive} [Gu: #1]}}
  
\else
  \long\def\comments#1{}
  \long\def\todo#1{}
  \long\def\zp#1{}  
  \long\def\jiawei#1{}
  \long\def\lx#1{}
  \long\def\del#1{}
  \long\def\gu#1{}
\fi
\setlist[enumerate]{
  itemsep=3pt,
  topsep=3pt,
  parsep=0pt,
  partopsep=0pt,
}

\setlist[itemize]{
  itemsep=3pt,
  topsep=3pt,
  parsep=0pt,
  partopsep=0pt,
}

\begin{document}
\pagestyle{plain}

\title{\tool: Automatic Configuration Repair for Production Networks}



\author{Zhenrong Gu}
\affiliation{%
 \institution{Xi'an Jiaotong University}
 \city{Xi'an}
 \country{China}}
\email{zrgu@stu.xjtu.edu.cn}

\author{Peng Zhang}
\affiliation{%
 \institution{Xi'an Jiaotong University}
 \city{Xi'an}
 \country{China}}
\email{p-zhang@xjtu.edu.cn}

\author{Xing Feng}
\affiliation{%
 \institution{Xi'an Jiaotong University}
 \city{Xi'an}
 \country{China}}
\email{xingfeng@stu.xjtu.edu.cn}

\author{Xu Liu}
\affiliation{%
 \institution{Xi'an Jiaotong University}
 \city{Xi'an}
 \country{China}}
\email{x.liu.reason@outlook.com}






\begin{abstract}

Network configurations are prone to errors, which can lead to catastrophic service outages.
A tool that can achieve automatic configuration repair (ACR) is highly desired by operators.
Existing tools for ACR follow a \emph{semantics-driven approach}:
they model network semantics as a set of SMT constraints, and solve them for a location or fix of the error. 
Due to the complex semantics of networks, constructing and solving these constraints can be prohibitively expensive,
making these tools neither general nor scalable.
Inspired by automatic program repair (APR), 
we explore another direction, i.e., a \emph{syntax-driven approach}, 
which generates and validates syntactically-valid candidate updates without modeling program semantics,
often drawing on existing code in the same repository.
Following this direction, we propose \tool, a syntax-driven method for ACR.
It uses multiple iterations of a ``localize-fix-validate'' pipeline to search for repairs, 
and proves quite effective on configurations of our production network.
Specifically, we show that Astragalus can repair every incident in multiple sizes of a synthesized network, and 97.5\% of the incidents on a real network, both with 15 types of errors injected,
within an average time of 6.93 seconds. It has also provided valid repairs in under 6 minutes for 7 recent network incidents or undesired changes, in a real production network with O(1,000)$\sim$O(10,000) devices. 

\end{abstract}

\maketitle

\section{Introduction}


Network outages frequently make headlines 
due to misconfigurations, software bugs, and hardware failures \cite{enwiki:1221077563}.
Among them, misconfiguration is a major source of network outages~\cite{liu2018automatic,zeng2012automatic}.
In the network of a large cloud service provider (CSP, referred to as \company~later), misconfigurations account for 35.4\% of all incidents in the past 5 years.

To detect network misconfigurations, both the academia and industry have proposed a lot of network verifiers \cite{fogel2015general,zhang2020apkeep,zhang2022differential,abhashkumar2019tiramisu,gember2016fast,fayaz2016efficient,beckett2017general,ye2020accuracy}.
These tools can check whether network forwarding behaviors align with operator intent, and conform to properties like loop freedom, and blackhole freedom.


However, network verifiers can only \textit{detect} issues or \textit{reduce} the impact rather than \textit{resolve} it,
and operators still need to manually localize the root causes and repair the configurations.
This manual process is frustrating and time-consuming, thereby delaying planned network updates; in \company, repairing an incident often takes far longer than detecting it (\S\ref{subsec:need}).


Similar to the Automatic Program Repair (APR) tools that greatly facilitate software development~\cite{gazzola2018automatic,monperrus2018automatic},
a tool that can achieve Automatic Configuration Repair (ACR) is highly desired by operators to help them quickly localize the root causes and resolve the incidents.

Towards ACR, there have been a handful of excellent tools \cite{gember2017automatically,abhashkumar2020aed,gember2022localizing}, which can localize configuration errors or generate repair for misconfigurations. 
These tools encode network semantics, forwarding properties, and configurations as SMT constraints and solve them with off-the-shelf solvers (e.g., Z3).
However, the complexity of distributed routing protocols limits both generality and scalability: constraints can only cover a subset of protocol features, and their count explodes with network size.
For a 45-node fat-tree running BGP, there are already 7900 SMT constraints—prohibitively expensive for solvers to handle in reasonable time.



For inspiration, we turn to APR, which has two broad classes of approaches: \textit{semantics-driven} approaches that solve SMT constraints for a correct-by-construction repair, and \textit{syntax-driven} approaches that \textit{generate} and \textit{validate} candidate \fix s without modeling program semantics (detailed in \S\ref{subsec:apr-inspiration}).
We observe that existing ACR methods, e.g., \cite{gember2017automatically,abhashkumar2020aed,gember2022localizing}, all fall within the first category of \emph{``semantics-driven'' approaches}, and thus inherit its scalability and generality limitations.

In this paper, we explore the ``syntax-driven'' approach and find it promising for ACR for two reasons.
First, the ``plastic surgery hypothesis'' holds for network configurations: devices of the same role carry similar configurations, providing the raw material for syntactically-valid fixes.
Second, unlike APR where validation dominates runtime due to recompilation~\cite{weimer2013leveraging}, network verifiers are fast enough to evaluate many candidate fixes efficiently.

Following the direction of syntax-driven approach, 
we propose \tool{}, a general and scalable tool that can achieve \acr~ for large-scale production networks.
\tool{} consists of three steps:
\textit{(1) Localization.}
We localize suspicious configuration lines with techniques from APR, Spectrum-Based Fault Localization (\sbfl)\cite{agarwal2014fault,jones2005empirical}.
The intuition for these techniques is that faulty lines contribute more to property violations than correct ones.
\textit{(2) Fix Generation.}
According to the suspiciousness of configuration lines, we generate a set of candidate fixes including remove, insert, and modify, 
based on existing configuration in the same network.
\textit{(3) Validation.} We use off-the-shelf configuration verifiers to 
check whether the \fix~resolves the violation without any side effects.





In summary, our contribution includes:
\begin{itemize}[leftmargin=*]
    \item We motivate a shift from semantics-driven paradigm to a syntax-driven paradigm for ACR, based on some insights from program repair.
    \item We design and implement \tool{} to realize \acr, a syntax-driven approach that is more scalable and general than existing methods based on SMT models.
    \item We test \tool{} on both real faults and synthesized network configurations with 15 types of injected errors, and show that it can repair all synthesized errors of O(100) real devices with an average time of 6.93 seconds.
\end{itemize}
\section{Motivation}


In this section, we first motivate the need for automatic configuration repair with real-world operational experiences (\S\ref{subsec:need}).
We then show that, despite a handful of prior efforts, existing approaches all fall into the \emph{semantics-driven} paradigm, which is correct but neither scalable nor general (\S\ref{subsec:semantics-driven}).
Finally, we draw inspiration from automatic program repair (APR), whose \emph{syntax-driven} paradigm motivates our approach (\S\ref{subsec:apr-inspiration}).

\subsection{The Need for Automatic Configuration Repair}\label{subsec:need}

\revision{
Network verification has matured: modern verifiers like \cite{fogel2015general} can check a large network against operator intent within minutes and are now routinely deployed in production.
As a result, \emph{detecting} a misconfiguration is rarely the bottleneck, but \emph{repairing} it is.
In \company, verification mostly finishes within $3$ minutes, yet operators still take more than $10$ minutes to localize and repair $45.8\%$ of incidents, with the worst case taking $5$ hours.
}

\revision{
Manual repair is slow because the causal chain from a configuration line to an intent violation is hard to trace.
Large networks run multiple routing protocols that interact over several rounds to compute routes; devices come from different vendors, each realizing features with vendor-specific behaviors; and multiple generations of network architecture often coexist, glued together by ad-hoc configurations to interoperate~\cite{gao2024crescent}.
Consequently, the relationship between a misconfiguration and the intent it violates is far from obvious, and even an experienced operator struggles to pinpoint the offending lines and craft a fix that restores intent without introducing new violations.
}

\revision{
This is the gap that motivates \emph{automatic configuration repair} (ACR): a tool that automatically localizes the root cause and produces a fix, much as Automatic Program Repair (APR) tools facilitate software development~\cite{gazzola2018automatic,monperrus2018automatic}.
}

\subsection{Limitations of Existing Approaches}\label{subsec:semantics-driven}

Towards ACR, there have been a handful of excellent tools that can localize configuration errors or generate repairs for misconfigurations.
\revision{Although they differ in goals and techniques, we observe that they all follow a \emph{semantics-driven} (a.k.a.\ correct-by-construction) recipe: they build a formal model of the network control plane and encode the repair problem as a set of SMT constraints, which an off-the-shelf solver (e.g., Z3) then solves.}

CPR~\cite{gember2017automatically} and AED~\cite{abhashkumar2020aed} encode the intent (e.g., blackhole-freedom) and the semantics of routing protocols as a set of SMT constraints, and introduce free variables for each line to represent candidate modifications.
An assignment of these free variables that makes the constraints satisfied is a repair.
While this SMT-based approach guarantees correct \fix s, it is hard to scale to large networks, due to the vast search space generated by SMT constraints and free variables.
Meanwhile, the repairs tend to focus on ``the most straightforward modifications'' that make the constraints satisfiable, which do not necessarily correspond to a true root-cause repair.
For example, on our dataset with injected misconfigurations, AED produced a repair that restores reachability by adding a static route.
Although this kind of ``workaround'' is functionally effective, it neither resolves the underlying configuration error nor provides a clear explanation of the root cause.

CEL~\cite{gember2022localizing} extends Minesweeper~\cite{beckett2017general} by decoupling configuration from the control logic and encoding the network model and intent using SMT constraints.
It localizes misconfigurations by solving for a Minimum Correction Set (MCS) of constraints, which maps back to the root-cause configuration statements.
However, since it targets fault localization, it cannot produce an actionable repair, and its scalability is similarly limited by constraint complexity and solver performance in large networks.

\revision{The root of these limitations is that network semantics, especially distributed routing protocols, are very complex.
As a result, the constraints can only cover a subset of protocols or features.
Worse, the number of constraints explodes with the network size and configuration complexity: for a fat-tree network of 45 nodes running BGP, the number of SMT constraints is already as large as 7900, which makes it prohibitively expensive for SMT solvers to solve in a reasonable time.}

\subsection{Inspiration from Automatic Program Repair}\label{subsec:apr-inspiration}

APR has been extensively studied, and its methods are broadly classified into two categories.
\textbf{Semantics-driven} approaches, also known as correct-by-construction approaches, formally encode the program repair problem as a satisfiability problem and solve it using off-the-shelf SMT solvers~\cite{nguyen2013semfix,mechtaev2015directfix}.
They guarantee the repair introduces no side effects, but scalability is limited because the entire problem can have many constraints.
\textbf{Syntax-driven} approaches, also known as generate-and-validate (G\&V) approaches, instead \textit{generate} candidate \fix s via syntactic modifications and \textit{validate} each against a test suite, without modeling program semantics~\cite{le2011genprog,liu2013r2fix,wei2010automated}.
Many build on the \emph{``plastic surgery hypothesis''}~\cite{barr2014plastic}: the code ingredients that can repair the software already exist somewhere in the codebase.
This makes G\&V more scalable by sidestepping expensive global semantic reasoning, and more general since no precise semantic model is needed.

\revision{\para{From APR to ACR.} ACR and APR share a highly similar problem structure: both start from an erroneous behavior and apply small, safe changes, filtered and validated against a consistent statement of intent (e.g. test cases), to drive the system from an incorrect state to one that satisfies the expected behavior.
The analogy runs deeper at the artifact level: a routing configuration, especially the route policies that steer routing protocols, reads much like a \emph{declarative program} -- a set of match-action rules that the control plane evaluates to produce routes.
This parallel suggests that the syntax-driven paradigm, which most existing ACR tools have overlooked, is worth exploring for configuration repair.}

\revision{\para{Unique challenges.} Network configuration repair is nonetheless not a direct transplant of APR, because configuration differs from a conventional program in two key ways.
First, its logic is not composed in the configuration alone but emerges from configuration coupled with protocol semantics and topology, where routes are computed by distributed protocols interacting over multiple rounds across devices, so modifying a single statement may affect the entire network, and a repair cannot be validated locally on a single device.
Second, the intent against which a repair is validated is often only vaguely defined: unlike the concrete test suite available in APR, operator intent is frequently implicit or partially specified, making it harder to state precisely.}

\section{Overview}
\label{sec_design}

\subsection{Key Ideas}
\label{subsec_idea}

The key idea of this paper is: 
instead of following the semantics-driven approach,
we explore the syntax-driven approach to achieve \acr~ for better scalability and generality.
This feasibility of applying syntax-driven approaches is due to the following two observations.


\para{The ``plastic surgery hypothesis'' holds true for network configurations.}
As observed in \company's network, 
the devices are organized into availability zones, planes and pods;
each of the zones also has multiple devices with the same role (i.e. core, aggregation, edge, etc.) to serve the purpose of redundancy.
The devices serving as each other's backup are often configured similarly;
and even the devices with the same role in different zones share similar configurations to achieve the same functionality~\cite{singh2015jupiter,facebook2014data_center_fabric}.
Therefore, the ingredients needed to generate syntactically-valid candidate fixes already exist in the network.
In practice, all seven real-world incidents we encountered can be repaired this way: each fix touches at most three configuration lines, with the necessary values already present in same-role devices, evidence that the ingredients needed for the repair already exist within the network (\S\ref{subsec:incident-a} and \S\ref{subsec:incident-b} give two examples).

\para{Validation is efficient for network configurations.}
Network verification is well-developed \cite{zhang2020apkeep,zhang2022differential} and can quickly check configuration correctness for large networks within seconds.
\company~has an in-house control plane simulator, which can finish the computation of the data plane for a production network with 2000 nodes, within 5 seconds, enabling us to efficiently try a lot of candidate fixes.
Note that in APR, the validation process dominates runtime due to the time required for recompiling and rerunning parts or the entire program~\cite{weimer2013leveraging}. 
Thus, the efficiency of validation implies the efficiency of repair for network configurations.

%
%





\begin{figure}
    \centering
    \includegraphics[width=\linewidth]{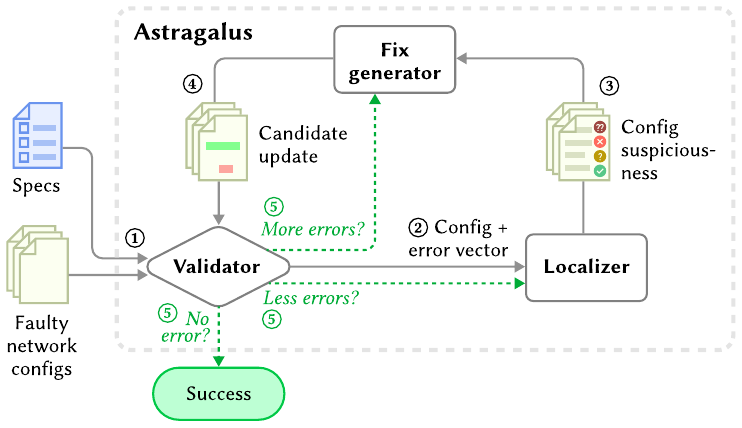}
    \vspace{-1.5em}
    \caption{The workflow of \tool{}. }
    \label{fig_workflow}
    \vspace{-2em}
\end{figure}

\subsection{Workflow}

Figure \ref{fig_workflow} shows a simplified workflow of \tool.
First, \tool~does a full validation of the network to identify the incident and its impact range. After that, \tool~tries to fix the network through multiple iterations of three steps: localization, fix generation, and validation.
 
\para{Step 1: Localization (\S\ref{sec:loc}).}

The localizer aims to identify the suspicious configurations that would cause the incident.
It tries to localize the fault based on the output from the validator.
With no extra information attached, it does the localization by best effort:
the identified configurations by the localizer are usually \textit{suspicious configurations} rather than the \textit{root causes}.
After the localization, the fix generator receives the suspiciousness of each part of the configuration.



\para{Step 2: Fix generation (\S\ref{sec_fix}).}

Based on the suspiciousness provided by the localizer, the fix generator generates \textit{candidate updates} heuristically, aiming to fix the suspicious config to resolve the incident.
Like the G\&V approach in APR, the fix generator selects one of the suspicious configuration and applies \emph{change operators} to produce a new configuration, i.e. the candidate update.

The change operators are purely syntactic, such as removing nodes in the abstract syntax tree (AST) of the configuration~\cite{baxter1998clone}, modifying a node or inserting a new node with values from other ASTs.

\para{Step 3: Validation (\S\ref{sec:validate}).}
After applying the change operator, the validator checks if the update satisfies the specification.
Since network verifiers~\cite{zhang2022differential,fogel2015general} can simulate the control plane from configurations, we can use off-the-shelf tools to validate candidate updates.

If the validator shows the network is fully fixed, it returns success. Otherwise, if it shows the network has fewer errors than before, it localizes again based on the new configuration (i.e., with the candidate update) to find new suspicious configurations;
if it shows the network is not improved, it tells the fix generator to try again by generating a new candidate update and validate it.

The above process iterates for several rounds until a repair is found,
or a maximum number of iterations is reached.



\subsection{Merits of \tool} 

The localize-fix-validate framework has three merits, compared with semantic-based approaches.

\begin{itemize}
\item \textbf{Scalability.} The search space for fix is relatively small, 
as it only involves a small subset of configurations related to the suspicious one.
\item \textbf{Generality.} The framework generalizes to different protocols and vendors, since it operates upon configuration syntax, rather than semantic. 
\item \textbf{Flexibility.} The framework allows a flexible choice of various techniques for each stage. For example, many \sbfl~methods can be applied for localization.
\end{itemize}

\section{Design}
\label{sec_design_new}

This section presents the design of \tool{}, which consists of three steps, i.e., localization, fix generation, and validation.



\subsection{Modeling the configuration}
\label{sec_int_sim}




As a syntax-driven approach to ACR, the prerequisite for \tool~to generate a valid fix is to keep the fix syntactically correct. This section focuses on the modeling of the configuration used by \tool.

\begin{figure}[ht!]
    \centering
    \includegraphics[width=\linewidth]{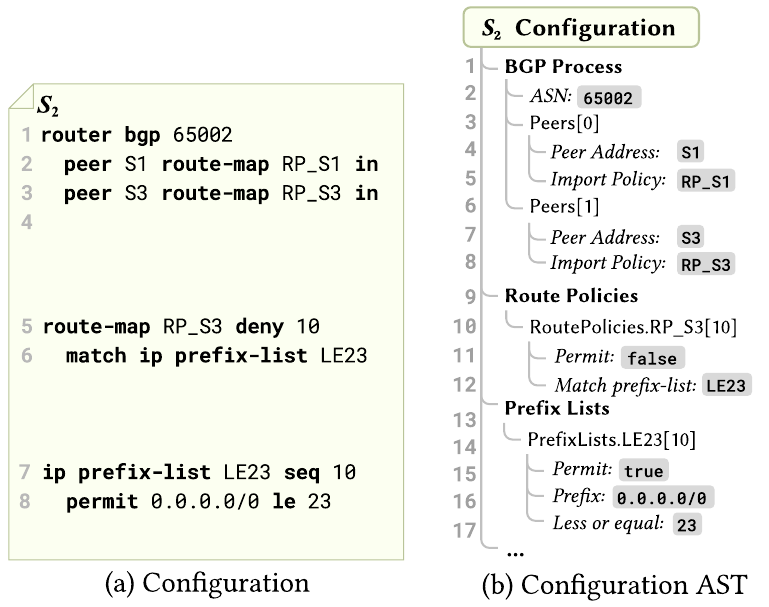}
    \vspace{-1.5em}
    \caption{A sample configuration snippet (a), the corresponding AST (b).}
    \label{fig_ast}
    \vspace{-1.5em}
\end{figure}

Like most routing simulators~\cite{fogel2015general}, \tool~represents configurations as an AST.
Figure \ref{fig_ast}(b) shows the AST for Figure \ref{fig_ast}(a): the tree organizes the configuration into BGP-process, route-policy, and prefix-list nodes, with leaves being atomic parameters (strings, IP addresses).
AST nodes are typed to enforce syntactic validity (e.g., a route policy node cannot be a child of a prefix list), and only some nodes support adding or removing children of their own type.

\tool~generates syntactically valid candidate updates on the AST.
To respect these restrictions, we define:
\begin{enumerate}
    \item A \textbf{configuration unit} is any node in the AST (used interchangeably with ``AST node'' hereafter).
    \item A \textbf{container configuration unit} is a node that supports adding or removing children of the same type, such as BGP peers, route policies, and prefix list rules.
\end{enumerate}

\begin{figure*}[ht!]
    \centering
    \includegraphics[width=\linewidth]{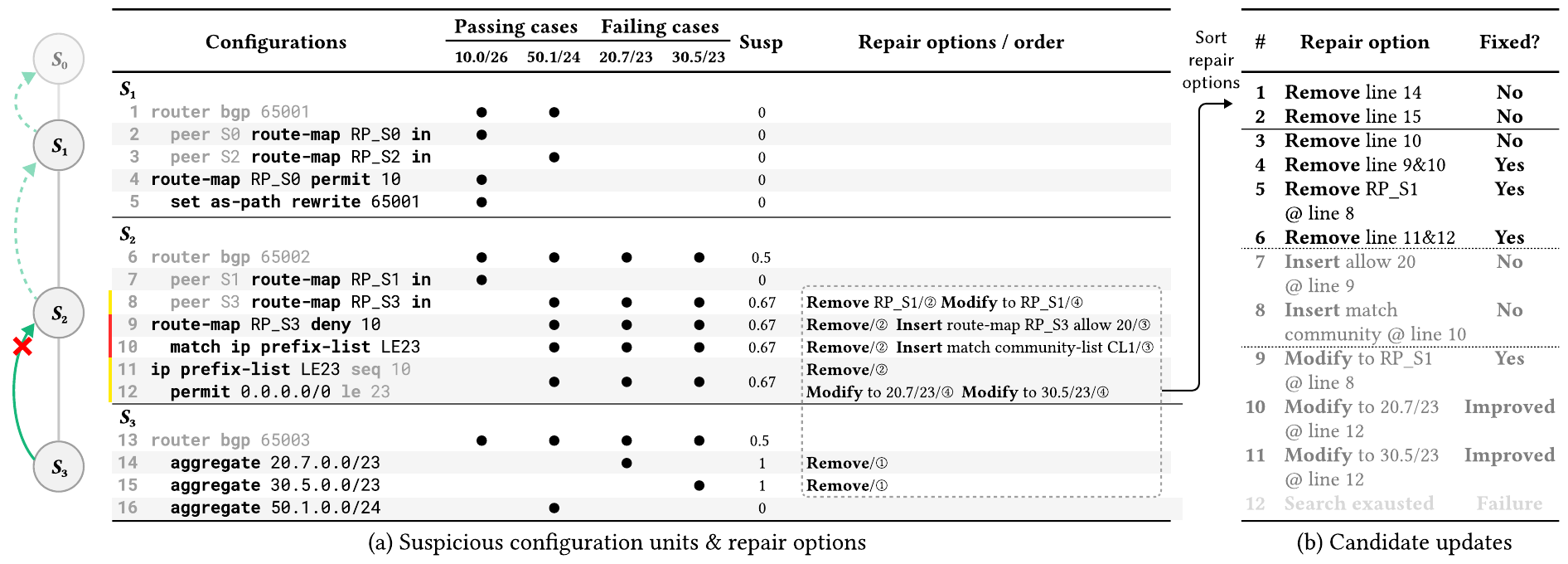}
    \caption{An example for localization and fix generation of \acr.
    The leftmost graph is an illustration of the topology.
    (a) The coverage report and suspiciousness calculation of each configuration unit in $S_1$, $S_2$ and $S_3$, using Tarantula; the repair options of each suspicious configuration unit.
    (b) The candidate repair options, and whether the repair option would fix the network. ``Improved'' means although the network still have issue, some of the failing prefix is solved.
    }
    \label{fig_walkthtrough}
\end{figure*}

With the terms and configuration model being defined, we will walk through the three key steps of \tool~in the following sections by demonstrating a repair example.
Figure \ref{fig_walkthtrough} demonstrates a typical configuration failure scenario.
In this topology, $S_3$ is configured with three BGP aggregate addresses 20.7.0.0/23, 30.5.0.0/23, and 50.1.0.0/24.
$S_0$ (the configuration is not shown) has an advertised network of 10.0.0.0/26.
Ideally, $S_2$ should propagate all three aggregate addresses from $S_3$ to $S_1$, but due to a misconfiguration in the import policy from $S_3$ (in line 9, it should \texttt{permit} rather than \texttt{deny} the routes), $S_2$ no longer receives the route, thus causing a black hole on prefix 20.7.0.0/23 and 30.5.0.0/23.

\subsection{Localization}
\label{sec:loc}




In APR, spectrum-based faults localization (SBFL) is a widely-used technique which generates suspiciousness scores for each line of a given program. 
Specifically, it uses a test suite \cite{abreu2007accuracy}, and generates an \textit{execution profile}, which contains:
\begin{enumerate}
    \item A \textit{coverage report}: what test cases are executed by each line of code.
    \item An \textit{error vector}: the pass / fail result for each test case.
\end{enumerate}

\tool~uses SBFL in the localization step.
The following section introduces how to construct the test suite, 
how to execute the test suite to form the execution profile, 
and how to compute suspiciousness of each configuration unit.

\para{Prefixes as the test suite.}
Viewing the control plane as a program, its inputs are prefixes announced inside the network (e.g., via \texttt{bgp network})\footnote{\texttt{bgp aggregate} prefixes are excluded because they are announced only when a more-specific route exists; we trace the more-specific route instead.} or advertised by external neighbors.
Unlike APR where the input space is vast, a network's prefix set is fully enumerable from device configurations.
\tool~therefore uses all \texttt{bgp network} prefixes as the test suite, supplemented by any relevant external route prefixes.



\para{Computing the configuration coverage report.}
Compared with APR, computing how each configuration unit is covered by a prefix seems to be tricky, since 
the routing and forwarding behaviors are jointly determined by the configurations, the control plane semantics, and the routing software / simulator implementation.

However, we can still see the configurations as executable ``code'':
When the routing software / simulator references one configuration unit while it is dealing with some route, we say the configuration unit is ``executed'', and the unit is considered covered by the prefix of the route.

For example, in Figure \ref{fig_walkthtrough}, propagating aggregate route 20.7.0.0/23 covers configuration units in sequence as the route travels: line 14 ($S_3$ aggregates the route) and line 13 ($S_3$ exports it to $S_2$); then, on $S_2$, line 6 (receiving the route), line 8 (matching the neighbor configuration for the incoming route), line 9 (applying import policy \texttt{RP\_S3}), line 10 (the policy's first node processes it), and lines 11--12 (testing the match condition \texttt{LE23}).

For \tool, we implement the coverage test functionality as an interface of the in-house simulator.
We start an iteration of data plane computation. The simulator records the coverage during its simulation process, and outputs the report to \tool.

The execution is initialized by simulating a single iteration of route advertisement from a converged data plane: each router sends all of its routes in the main RIB to neighbors.
The insight of starting from a converged data plane is as follows:
(1) the test cases (i.e. the \texttt{bgp network}s' and external routes' prefixes) are propagated to every destination as the control plane program suggests, so every logic that should be covered by the test cases is covered;
(2) because of the convergence, route withdrawal in transient states will not affect the coverage.
The converged data plane can be derived from the validation step of the previous \acr{} iteration to avoid duplicate calculation.
By aligning \tool's AST with that of the simulator's, the coverage report can be used seamlessly by \tool.

\para{Generating the error vector.}
In \tool, we use \textit{specifications}, which are specified by the operators, to check whether a test case passes or not.
We use per-prefix forwarding graph as the default specification of \tool, as it implies many invariants in a network, such as reachability, blackhole freedom, loop freedom, route-hijacking freedom, and ECMP multipath. The forwarding graph can also be automatically generated in a well-formed hierarchical network like fat-trees.
For each prefix in the test suite, \tool~generates the forwarding graph of that prefix from the data plane computed by the simulator, and checks if the forwarding graph is different from the one in the specification.
If the path violates the specification, \tool~adds the prefix to the set of \emph{failing tests};
otherwise, \tool~adds it to the set of \emph{passing tests}.

\tool~enables a wide range of specifications choices, as long as a specification can test whether a prefix meets its requirement.
For example, all-pair reachability is a parameter-free specification which can be adapted by \tool; the specification reports failure if and only if any of the nodes cannot reach the owner of the prefix.

In the example of Figure \ref{fig_walkthtrough}, we mark two prefixes (10.0.0.0/26, 50.1.0.0/24) as passed, as they can be reached from all other devices;
we mark the other two prefixes (20.7.0.0/23, 30.5.0.0/23) which are denied by line 9 and 10 are marked as failed.

\label{sec_loc}

\para{Calculating suspiciousness scores.}
\tool~can use any SBFL technique to calculate suspiciousness scores. Among the well-known methods like Tarantula \cite{jones2005empirical}, Ochiai \cite{abreu2007accuracy}, Jaccard \cite{chen2002pinpoint}, and D-Star\cite{wong2012software}, \tool~uses Ochiai by default, 
to calculate the suspiciousness scores for configuration units.
According to Ochiai, the suspiciousness of a statement $s$ is calculated as:
\begin{equation}\label{equ_tarantula}
    \mathrm{susp}(s) = \frac{\mathrm{failed}(s)}{\sqrt{(\mathrm{failed}(s) + \mathrm{passed}(s))\cdot \mathrm{totalfailed}}}
\end{equation}
where $\mathrm{failed}(s)$ and $\mathrm{passed}(s)$ represent the number of failed and passed test cases executed by the statement $s$, respectively;
Meanwhile, $\mathrm{totalpassed}$ and $\mathrm{totalfailed}$ represent the total number of passed and failed test cases, respectively.
The decision of using Ochiai SBFL technique as default is based on the evaluations in \S\ref{sec:loc-quality}.

\subsection{Generating the Fixes}

\label{sec_fix}
The fix generation step generates candidate repairs for the suspicious configuration units computed by the localization step.


\para{How to fix.}
For each suspicious configuration unit, \tool~ can use three \emph{change operators} to fix the suspicious configuration unit.
\begin{itemize}
\item \textit{Remove} deletes the node (and its subtree) of the suspicious configuration unit from the AST.
\item \textit{Insert} inserts a node into the suspicious configuration unit in the AST.
\item \textit{Modify} changes the value of the suspicious configuration unit to other values different from the current one.
\end{itemize}


\para{Where to apply \textit{Remove} / \textit{Insert} / \textit{Modify}.}
As \S\ref{sec_int_sim} establishes, only \textbf{container configuration units} support child insertion and deletion, so \textit{Remove} and \textit{Insert} apply only to containers; \textit{Modify} applies to any configuration unit.


\para{What to \textit{Insert} / \textit{Modify}.} As the plastic surgery hypothesis suggests, \tool{} inserts into / modifies the suspicious configuration unit with nodes that exists elsewhere.
At first glance, it seems there are a large number choices for configuration lines to copy and parameters values to modify.
However, we find that the number of all feasible choices is quite limited for network configurations, due to device roles and configuration syntax.
More specifically, \tool{} selects a device of the \emph{same role} (including the device itself) as the template device,
and chooses a configuration unit of the \emph{same type} as the suspicious unit.

Figure \ref{fig_walkthtrough}(a) shows the fixes that can be generated from the suspicious configurations, only part of the \textit{Modify} and \textit{Insert} options are shown. For example, line 8 shows the possible repair options for configuration unit \texttt{route-map RP\_S3 in}. As the import policy is optional (i.e. a container configuration unit with capacity of 1), the \textit{Remove} option can be applied; it may also paste the route policy name \texttt{RP\_S1} from line 7.


\para{Deciding the order of fixes.}
\tool{} generates all possible fixes for configuration units with a suspiciousness score $> 0$.
Then, \tool{} sorts the possible fixes from high to low based on the suspiciousness of the configuration unit from which each fix is generated.
If multiple configuration units have the same score, then \textit{Remove} updates are placed higher in the list and searched earlier, since there are at most one removing option per configuration unit, and it is more likely to yield a successful repair based on our tests;
\textit{Insert} updates are placed after \textit{Remove}, and finally \textit{Modify}; if the comparison is a tie, their order in the list is determined randomly.

The order is decided based on the heuristics of the data center studied.

The list then forms all candidate updates of the network, as shown in Figure \ref{fig_walkthtrough}(b).

\subsection{Validating the Fixes}
\label{sec:validate}

For each fix, \tool{} first applies the fix to the configurations, thus forming a new control plane consisting of the new configurations.
Then \tool{} computes the data plane, and validate the fix by testing if all test cases are passed, using the specification described by \S\ref{sec_loc}.


The above process of generate-and-validate is repeated for multiple iterations until a repair is found.
In each iteration,
we apply the topmost repair option in the candidate update list (as shown in Figure \ref{fig_walkthtrough}) to our network, 
and a new set of configurations with the repair applied (e.g. the removal of line 6 in $S_2$) is produced.
By viewing the old and new set of configurations as two states, 
each iteration can be seen as transition from the old state to the new one.
Therefore, \tool{} needs a search strategy to efficiently explore the state space of configurations.

Currently, \tool{} adopts a depth-first strategy as follows.
After each iteration, 
if the new configuration state produces less errors, 
then we fork a new iteration based on the new state,
and start searching on the state immediately.
To avoid duplicates, we will check whether the state has been visited before executing the iteration.
We also backtrack to the previous state if the repair option is exhausted.

If we find the state that produce no error, we report a success, with the path to the final state as the final repair.

\section{Evaluation}

In this section, we evaluate \tool~to answer the following research questions:
\textbf{(RQ1)} Can \tool~accelerate the repair / fault localization of network configurations?
\textbf{(RQ2)} Can \tool~improve the scalability of ACR to larger production networks?
\textbf{(RQ3)} Can \tool~provide higher quality repairs than existing tools?
\textbf{(RQ4)} Is localization (SBFL) effective for repairing network configurations?

\subsection{Implementation}

We implemented \tool{} with $\sim 10$K lines of Rust code. \tool~is implemented with two parts:
\begin{itemize}
    \item ACR-SDK is an SDK that provides utilities for Rust-based simulators to meet all the requirement to implement \acr{} on top: it provides utilities to insert hooks and generate coverage report in its simulation logic; it also provides functions for localization and fix-generation.
    \item \tool~is developed as a feature of the in-house simulator that uses ACR-SDK to fully implement ACR.
\end{itemize}



\subsection{Dataset and Setup}

\para{Real dataset.}
The real dataset is based on a subset of the production network of the CSP we studied.
It is sampled based on the principle that we should be able to reproduce all historical incidents.
If certain devices were taken offline due to architectural evolution, we manually selected substitute online devices with the same role.
In the end, the dataset consists of 83 devices with 306 lines of configuration on average, preserving the topology and configuration of the production network.


\para{Synthetic dataset
.
}
We also synthesized a dataset of 5 different sizes ($k=4, 6, 8, 10, 12$) of fat-tree topology \cite{al2008scalable} based on \cite{raghunathan2022acorn}.
Each device has 97.4 $\sim$ 163 lines of configuration on average, respectively.
To reflect the characteristics of the CSP's real data center networks (DCNs), we have modified the configurations as follows.
\begin{itemize}
    \item In DCNs, the devices at the same layer reuse the ASNs.
    As a result, the AS-path of a BGP route sent across a layer twice may contain two identical ASNs.
    To prevent routes from being dropped because of the eBGP loop prevention mechanism,
    we configure \texttt{as-path rewrite} on the aggregation devices, and the corresponding route policies.
    \item In DCNs, devices configure BGP aggregation to reduce the total number of routes.
    We configure BGP aggregation on the aggregation devices, 
    and route policies on the core devices to only import aggregate routes.
    \item DCNs are constantly expanding, leaving some legacy configurations in place.
    Even if these legacy configurations are not used, they are still kept to avoid unexpected incidents, causing redundancy in configurations.
    In the route policies created above, we mimic this characteristic by introducing some redundant route policies, e.g.,
    a route can be matched by a community list, a prefix list, or both.
\end{itemize}

\begin{table}[ht!]
    \centering
    \footnotesize
    \caption{
    Misconfigurations and support of tools.
    }
    \vspace{-1.0em}
    \begin{tabular}{l l c c c}
    \toprule
    \multirow{2}{*}{\textbf{ID}} & \multirow{2}{*}{\textbf{Description}} & \multicolumn{3}{c}{\textbf{Supported by}} \\
    \cmidrule(lr){3-5}
    & & AED & CEL & \tool\\
    \midrule
    1    & Wrong set as-path (shorter)        &              &              & $\checkmark$ \\
    2    & Wrong route policy application     & $\checkmark$ & $\checkmark$ & $\checkmark$ \\
    3    & Missing prefix list rule           & $\checkmark$ & $\checkmark$ & $\checkmark$ \\
    4    & Missing route-policy node          & $\checkmark$ & $\checkmark$ & $\checkmark$ \\
    5    & Additional aggregate address       &              &              & $\checkmark$ \\
    6    & Additional route policy node       & $\checkmark$ & $\checkmark$ & $\checkmark$ \\
    7    & Additional prefix list rule        & $\checkmark$ & $\checkmark$ & $\checkmark$ \\
    8    & Wrong route policy node permit     & $\checkmark$ & $\checkmark$ & $\checkmark$ \\
    9    & Wrong prefix-list rule permit      & $\checkmark$ & $\checkmark$ & $\checkmark$ \\
    10   & Wrong match-prefix                 & $\checkmark$ & $\checkmark$ & $\checkmark$ \\
    11   & Wrong route policy node priority   & $\checkmark$ & $\checkmark$ & $\checkmark$ \\
    12   & Wrong set as-path (longer)         &              &              & $\checkmark$ \\
    13   & Missing route policy application   & $\checkmark$ & $\checkmark$ & $\checkmark$ \\
    14   & Missing set as-path                &              &              & $\checkmark$ \\
    15   & Additional set as-path             &              &              & $\checkmark$ \\
    \bottomrule
    \end{tabular}
    \vspace{-1.0em}
    \label{tab:misconf}
\end{table}

\para{Misconfigurations.}
We consider 15 types of configuration errors, as shown in Table \ref{tab:misconf}. 
The first 7 are real historical bugs, while the remaining 8 are bugs that we hypothesized to be at risk of occurring.
For each type of error, we generate 100 instances for the above two configuration datasets. 
Specifically, we generate \textit{missing} errors by randomly removing a corresponding configuration,
\textit{additional} errors by copying and pasting a corresponding configuration from a random device, 
and \textit{wrong} errors by changing a parameter to some value, which can be a flip of ``permit''/``deny'', or a value pasted elsewhere.

These errors may lead to two types of incidents:
(1) reachability incidents, where one or more devices cannot reach a prefix;
and (2) path change incidents, where the forwarding paths in the misconfigured network are different from the original ones.
Note that some types of the misconfigurations may not cause a reachability incident in some networks. For instance, changes in the Aspath length will not affect the network we have generated.
For the real configuration dataset, 100 errors cause path change incidents for each error type.
For the synthesized configuration dataset, 50 errors cause path change incidents for each error type, 
and 50 errors cause reachability incidents only for ER2$\sim$ER11.

\para{Compared methods.} We compare \tool~ with two state-of-the-art methods -- AED \cite{abhashkumar2020aed} which can synthesize repairs to make the configurations satisfy specified properties,
and CEL \cite{gember2022localizing} which can localize the configuration errors that cause property violations,
but cannot return any repair.
The comparison is only done for errors causing reachability incidents, as the two methods do not support specifying forwarding paths.


\para{Setup.} The tests are configured with a timeout of 10000 seconds.
All experiments were run on a bare-metal server, with 2x AMD EPYC 7Y83 64 core CPU running at 3.1GHz and 1007GiB of memory. Each individual repair attempt was run in parallel, with 8 logical CPU-core assigned using Linux CPU affinity.

\subsection{(RQ1) Repair Performance}

\begin{table}[ht!]
    \centering
    \footnotesize
    \caption{
    Average time (in seconds) of repairing / locating errors causing two kinds of incident of AED, CEL and \tool~in fat-tree ($k=4$) in the public dataset. ``$\times$'' means the error is not supported by the tool, and ``-'' marks the unavailable dataset / time.
    }
    \vspace{-1.0em}
    \begin{tabular}{c c c c c c c}
    \toprule
    \multirow{2}{*}{\textbf{Error ID}} & \multicolumn{3}{c}{\textbf{Reachability (s)}} & \multicolumn{3}{c}{\textbf{Path Change (s) }} \\
    \cmidrule(lr){2-4}\cmidrule(lr){5-7}
    & AED & CEL & \tool & AED & CEL & \tool \\
    \midrule
    1    & -        & -        & -     & $\times$ & $\times$ & 0.291 \\
    2    & 77.59    & 86.18    & \textbf{0.019} & $\times$ & $\times$ & 0.054 \\
    3    & 67.11    & 86.27    & \textbf{0.016} & $\times$ & $\times$ & 0.045 \\
    4    & 66.90    & 85.70    & \textbf{0.029} & $\times$ & $\times$ & 0.245 \\
    5    & $\times$ & $\times$ & \textbf{0.066} & $\times$ & $\times$ & 0.067 \\
    6    & 76.59    & 105.51   & \textbf{0.017} & $\times$ & $\times$ & 0.066 \\
    7    & 112.28   & 87.63    & \textbf{0.015} & $\times$ & $\times$ & 0.073 \\
    8    & 112.46   & 83.94    & \textbf{0.027} & $\times$ & $\times$ & 0.093 \\
    9    & 111.84   & 141.73   & \textbf{0.016} & $\times$ & $\times$ & 0.046 \\
    10   & 109.15   & 84.72    & \textbf{0.018} & $\times$ & $\times$ & 0.071 \\
    11   & 108.79   & 84.98    & \textbf{0.013} & $\times$ & $\times$ & 0.052 \\
    12   & -        & -        & -              & $\times$ & $\times$ & 0.065 \\
    13   & -        & -        & -              & $\times$ & $\times$ & 0.156 \\
    14   & -        & -        & -              & $\times$ & $\times$ & 0.057 \\
    15   & -        & -        & -              & $\times$ & $\times$ & 0.072 \\
    \midrule
    Average & 93.63 & 94.30    & \textbf{0.024} & -        & -        & 0.097 \\
    \bottomrule
    \end{tabular}
    \vspace{-1.0em}
    \label{tab:public-time}
\end{table}

\para{Performance on the synthetic dataset.}
For the synthetic dataset, all three tools can correctly repair / locate 100\% of the faults.
Table \ref{tab:public-time} shows the time of repairing (for AED and \tool) / locating (for CEL) the misconfiguration that causes the error.
As shown on the table, in a relatively small network, \tool~is able to repair the fault 4$\sim$5 orders-of-magnitude faster than the state-of-the-art solutions like AED and CEL.

Both AED and CEL cannot repair / locate Error 5, because in minesweeper, once a device is identified as the destination device, it will unconditionally advertise the dst IP to its neighbors, effectively ignoring the impact of aggregate routes; therefore, even if extra aggregate routes are injected into the edge device for ER5, they do not change the original reachability, and AED and CEL will not treat this as an error. 


\begin{table}[ht!]
    \centering
    \footnotesize
    \caption{
    Success rate and time of repairing errors in the real dataset.
    }
    \vspace{-1.0em}
    \begin{tabular}{c c c c}
    \toprule
    \textbf{Error ID} & \textbf{Success Rate} & \textbf{Median Time (s)} & \textbf{Average Time (s)}\\
    \midrule
    1    & 91\%  & 0.38  & 1.02 \\
    2    & 100\% & 10.06 & 12.57 \\
    3    & 97\%  & 0.23  & 1.63 \\
    4    & 98\%  & 5.89  & 9.82 \\
    5    & 100\% & 0.06  & 7.31 \\
    6    & 92\%  & 1.08  & 11.35 \\
    7    & 100\% & 0.14  & 2.31 \\
    8    & 100\% & 4.37  & 7.03 \\
    9    & 100\% & 0.25  & 0.27 \\
    10   & 98\%  & 0.22  & 3.11 \\
    11   & 100\% & 12.35 & 9.87 \\
    12   & 100\% & 0.49  & 0.70 \\
    13   & 87\%  & 0.05  & 16.39 \\
    14   & 100\% & 0.08  & 0.15 \\
    15   & 100\% & 4.75  & 20.41 \\
    \midrule
    Overall & 97.53\% & 0.41 & 6.93\\
    \bottomrule
    \end{tabular}
    \vspace{-1.0em}
    \label{tab:real-succ-time}
\end{table}

\para{Performance on the real dataset.}
Table \ref{tab:real-succ-time} shows the success rate and average repair time of \tool~in the real dataset. As shown on the table, in the real dataset with more devices and configurations, \tool~is still able to achieve an average repair time of 6.93 seconds, while most of the errors can be repaired in less than a second.

Notably, the success rate of \tool~is no longer 100\%.
The most significant reason is that in the reduced data center topology, there are devices with a role only appears once in the data center; and some configurations are unique across the network.
When these configurations are removed or altered, the plastic-surgery hypothesis no longer guarantees a valid fix.
However, when applying to a real data center network, this phenomenon could be ignored, as the devices usually have more redundancy.

\subsection{(RQ2) Repair Scalability}

\begin{figure}[ht!]
    \centering
    \vspace{-1.0em}
    \includegraphics[width=0.9\linewidth]{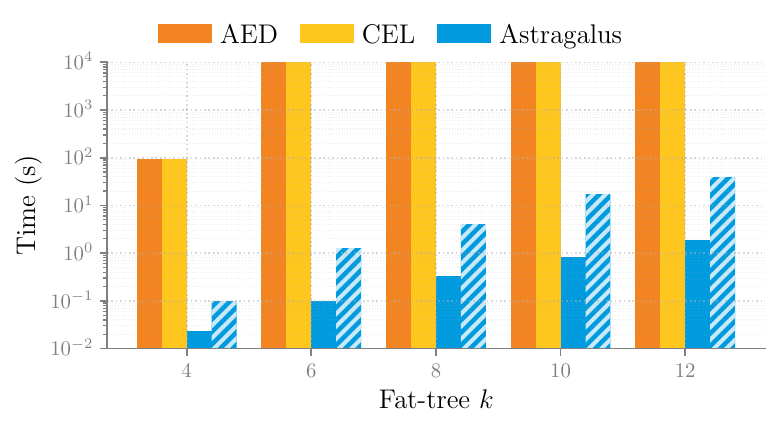}
    \vspace{-1.5em}
    \caption{The average repair / locate time of AED, CEL, and \tool.
    Solid bars is for reachability incidents, and stripped bars is for path change incidents.}
    \vspace{-1.5em}
    \label{fig:ft-levels}
\end{figure}

Figure \ref{fig:ft-levels} shows the repair time of \tool~over different sizes of fat-trees in the synthetic dataset, alongside AED and CEL.
The $x$-axis denotes the fat-tree size $k$ (ranging from 4 to 12), while the y-axis reports repair/location time in seconds on a logarithmic scale.
As evidenced in the figure, AED (orange bars) and CEL (yellow bars) exhibit severe scalability limitations.
For~$k = 4$, both tools require approximately 100 seconds to complete their tasks. However, for $k \geq 6$ , their execution times plateau at the 10,000-second timeout,
indicating they are no longer practical for moderately sized networks.
In contrast, \tool~(blue bars) demonstrates exceptional scalability, with repair times remaining orders of magnitude lower across all network sizes.
As network size increases to $k = 12$, \tool’s repair times rise gradually to less than 1 minute for path change incidents, and remain at 1.9 seconds for reachability incidents.

The difference in time of faults causing reachability incidents and path change incidents is also noticeable.
As fat-tree is a robust network topology for networks with redundancy in mind, there are fewer ways to introduce a fault by altering a single configuration unit,
making the error profile more monotonic.
Moreover, these errors are more likely injected into the ``weaker'' points in the network,
e.g. the edge switches' export policy to the aggregation layer.
As these weak points break, the routes of the faulty prefixes are only propagated to very few devices,
making the number of configuration units covered by these prefixes small;
further concentrating the selection range of SBFL.

\subsection{(RQ3) Repair Quality}

To quantify the repair quality of \tool, we evaluate how many repairs are produced by directly removing the root cause of the problem,
i.e. removing the misconfiguration being injected.
we define \textit{hit ratio} as the frequency of a tool giving a repair with the location \textit{exactly the same} as the location where the misconfiguration is injected.
That is, if a repair is ``add a static route'', it does not count as a hit since it does not localize the root cause.
This metric roughly estimates the probability that the repair solutions provided by \tool~are \textit{directly adopted} by operators.

For each network to be repaired, CEL gives multiple \textit{candidate predicates}, which can be seen as the possible locations of misconfigurations.
To apply a fix, network engineers also need to investigate all of the candidate predicates.
We calculate the frequency of the root cause as the frequency of the root cause in all candidate predicates.

\begin{table}[ht!]
    \centering
    \footnotesize
    \caption{
    Average hit ratio of AED, CEL and \tool~in fat-tree ($k=4$) in the synthetic dataset. ``$\times$'' means the tool cannot locate the error.
    }
    \vspace{-1.0em}
    \begin{tabular}{c c c c c }
    \toprule
    \multirow{2}{*}{\textbf{Error ID}} & \multicolumn{3}{c}{\textbf{Reachability}} & \textbf{Path Change} \\
    \cmidrule(lr){2-4}\cmidrule(lr){5-5}
    & AED & CEL & \tool & \tool \\
    \midrule
    1    & -        & -        & -              & 40\% \\
    2    & $\times$ & \textbf{50\%} & 12\%  & 18\% \\
    3    & $\times$ & \textbf{20\%}    & 0\%   & 0\% \\
    4    & $\times$ & 25\%    & \textbf{100\%} & 82\% \\
    5    & $\times$ & $\times$ & \textbf{100\%} & 100\% \\
    6    & $\times$ & 25\%   & \textbf{100\%} & 78\% \\
    7    & $\times$ & 17.36\%    & \textbf{64\%}  & 66\% \\
    8    & $\times$ & 25\%    & \textbf{100\%} & 82\% \\
    9    & $\times$ & \textbf{4.68\%}   & 0\%   & 0\% \\
    10   & $\times$ & \textbf{20\%}    & 0\%   & 16\% \\
    11   & $\times$ & 25\%    & \textbf{100\%} & 76\% \\
    12   & -        & -        & -              & 0\% \\
    13   & -        & -        & -              & 0\% \\
    14   & -        & -        & -              & 0\% \\
    15   & -        & -        & -              & 100\% \\
    \midrule
    All & 0\%         & 16.33\%   & \textbf{57.6\%} & 43.9\% \\
    \bottomrule
    \end{tabular}
    \vspace{-1.0em}
    \label{tab:hit-ratio}
\end{table}

Table \ref{tab:hit-ratio} presents the hit ratio of AED, CEL, and \tool.
A key note is that despite CEL’s lower overall hit ratio compared to \tool, it includes all root causes within its candidate predicates,
which means network engineers can still identify the true issue by investigating the full set of CEL’s suggestions.

For all test cases, AED exclusively proposes adding static routes, which do not localize the root causes,
resulting in a hit ratio of 0\% across all errors.
The hit ratio of \tool~highly relies on the misconfiguration type and the priority of the candidate updates.
If an easier approach (e.g. changing a node of route policy from ``deny'' to ``permit'') can fix the network,
\tool~would not search further for the root cause, e.g. adding a new route policy.
This leads to the vast difference of the hit ratio: errors with easier root causes tend to have higher hit ratio.

A low \textit{hit ratio} also does not necessarily means low fix quality. 
For instance, a fault introduced by removing a route policy node may be fixed by changing the route policy to another one.
Since the location of the fix is different from the injected misconfiguration,
it would not be counted as a ``hit'', but it would also reveal the root cause of the fault to the network operators. 

\subsection{(RQ4) Localization quality}
\label{sec:loc-quality}
\tool~searches candidate updates guided by the suspiciousness given by SBFL.
Without an efficient localization strategy, \tool~may end up with randomly searching and validating in space of all configuration units,
which is extremely time consuming.
As a result, the localization quality directly determines the efficiency and quality of \tool.

We define \textit{relative root cause location} $L_r$ as the ratio of position of the root cause configuration in the list to the number of all suspicious configuration units.
For example, if the suspiciousness of the root cause is 0.5, and there are 5 configuration units with suspiciousness higher than 0.5 out of a total of 100 configuration units, the relative root cause position is $5 / 100 = 0.05$.

The distribution of $L_r$ indicates how effective the localization is.
If we choose the fix randomly from all suspicious configuration units, $L_r$ should satisfy uniform distribution, where the root cause should be found anywhere with equal possibility density;
if the localization method is effective, we should expect more root causes to be found in prior positions.

\begin{figure}[ht!]
    \centering
    \vspace{-1em}
    \includegraphics[width=\linewidth]{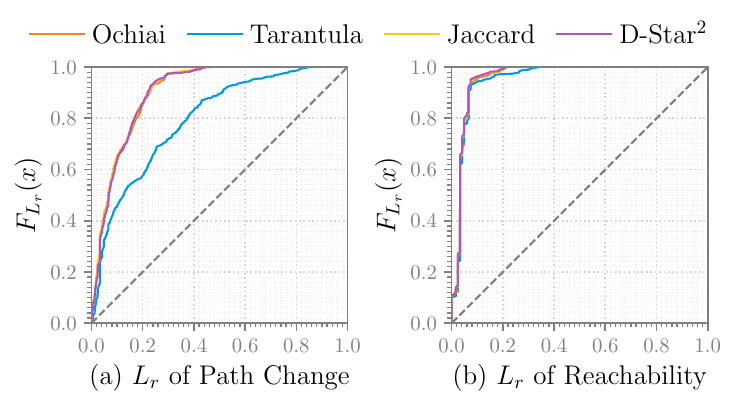}
    \vspace{-1em}
    \caption{CDF of \textit{relative root cause location} ($L_r$) of the 4 SBFL techniques. The diagonal line indicates $L_r$ of selecting suspicious configurations randomly.}
    \vspace{-1em}
    \label{fig:cdf}
\end{figure}

Figure \ref{fig:cdf} shows the cumulative distribution function (CDF) of the relative root cause location $L_r$ for 4 SBFL techniques: Ochiai, tarantula, jaccard, and D-Star (with parameter 2).
In both path change and reachability test cases, Ochiai, Jaccard and D-Star$^2$ do not have observable difference in the CDF;
but all of the three methods can find the root causes faster than Tarantula.

We choose Ochiai as the default because it is one of the three faster SBFL techniques, but using Jaccard and D-Star$^2$ will not make localization observably less accurate.
Choosing Tarantula, however, would make the localization, and the repair slower.
In fact, in fat-tree with $k$ as small as 6, Tarantula cannot find a valid repair within 6000 iterations for 9 cases of path change incidents, making the repairing time more than 5 minutes;
while using Ochiai, all faults in the synthesized dataset can be repaired within 2607 iterations.



\subsection{Microbenchmarks}

To gain a deeper understanding of the characteristics of the repairing process and outcomes by \tool,
we tested (1) the time of the three steps -- localization, fix generation and validation, taken separately;
(2) the proportion of three operators -- \textit{Remove}, \textit{Insert} and \textit{Modify}, used in the repair.

\label{app:mb-result}

\begin{figure}[ht!]
    \centering
    \vspace{-1.0em}
    \includegraphics[width=\linewidth]{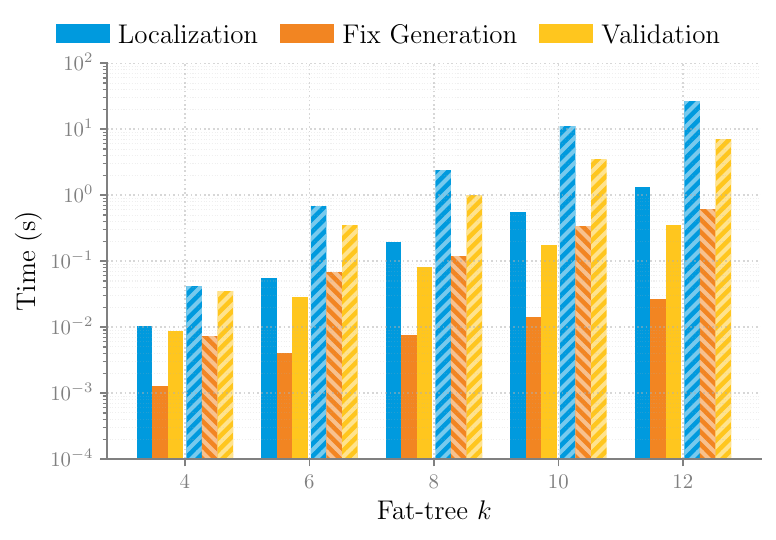}
    \vspace{-1.5em}
    \caption{The average time of localization, fix generation, and validation of \tool, in the synthesized dataset.
    Solid bars is for reachability incidents, and stripped bars is for path change incidents.}
    \vspace{-0.5em}
    \label{fig:lfv}
\end{figure}

Figure \ref{fig:lfv} shows the running time of the three modules, in the synthesized dataset.
We see that most time is spent on localization and validation, where the simulator involves to compute the data plane / generate the coverage report.
Generating fixes, however, takes much less time since the simulator is not involved.

\begin{figure}[ht!]
    \centering
    \vspace{-0.3em}
    \includegraphics[width=0.8\linewidth]{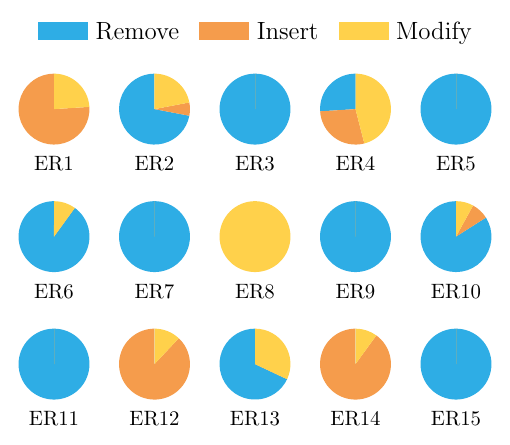}
    \vspace{-1em}
    \caption{The proportion of three change operators applied in fixing fat-tree 8's path change incidents.}
    \vspace{-1em}
    \label{fig:rim}
\end{figure}

Figure \ref{fig:rim} reveals the proportion of the three change operators for fixing each kind of misconfigurations.
In total, although ER5, ER6, ER7 and ER15 are the only 4 out of the 15 misconfigurations whose root cause should be removed,
66.5\% of the repairs use the \textit{Remove} change operator.
The proportion of \textit{Remove} operator is even higher for the reachability incident dataset:
79\% of the faults can be repaired with the \textit{Remove} operator.
In the real dataset, \textit{Remove} is still the most dominant change operator, which can fix 37.4\% of the faults.
The second most used operator is \textit{Insert}, counting up for 18.4\% of the fixes.
Only 15.1\% of the cases are fixed by \textit{Modify} operator.

\section{Real-World Experiences}

As a tool aiming for ACR in real production networks, \tool~has been tested on a large service provider's networks, 
which have O(1,000) $\sim$ O(10,000) devices.
In the following, we share 2 of the incidents highlighting the scalability, repair quality, and limitations of \tool.

\begin{figure}[ht!]
\vspace{-1em}
\includegraphics[width=\linewidth]{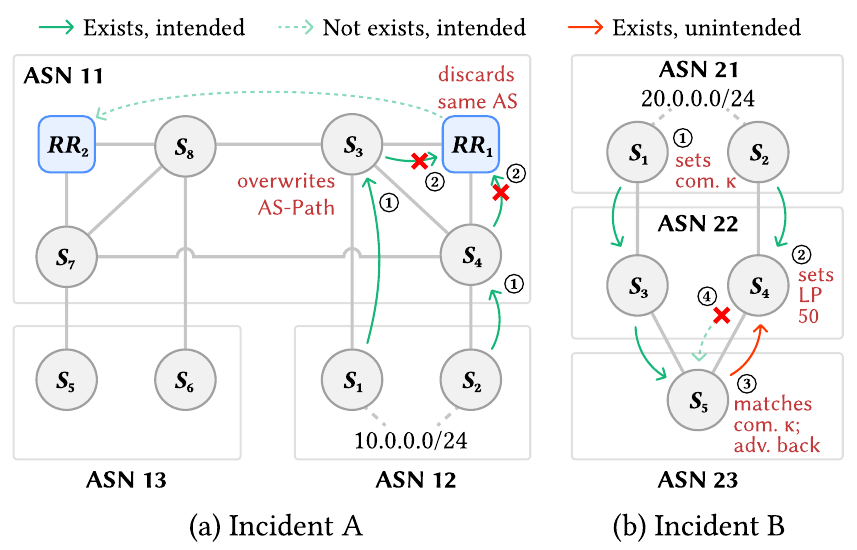}
\vspace{-1em}
\caption{Two real-world network incidents studied. Arrows indicate route propagation, and circled numbers mark the order of events. For Incident B, only $S_4$'s perspective is drawn to avoid clutter; the same flapping occurs symmetrically on $S_3$.}
\label{fig:case-study}
\vspace{-1.5em}
\end{figure}





\subsection{Incident A: Blackhole due to VSBs}\label{subsec:incident-a}

\para{Network and configurations.}
The network comprises $\sim$1000 devices, and each device has $\sim$20,000 lines of configuration on average.
As shown in Figure \ref{fig:case-study}(a), the network spans multiple data centers (only two are shown, AS12 and AS13), connected by a transit AS (AS11).
In AS11, each neighboring data center is served by two peering switches ($S_3$, $S_4$ for AS12; $S_7$, $S_8$ for AS13) and a BGP route-reflector (RR).
The two RRs peer over IBGP to exchange routes between the data centers, while each peering switch establishes IBGP only with its corresponding RR ($S_3, S_4 \leftrightarrow RR_1$; $S_7, S_8 \leftrightarrow RR_2$).

Many data centers in the network reuse ASNs in some layer of the network.
As a result, the AS-path of a BGP route sent from one data center to another may contain two identical ASNs.
To prevent routes from being dropped by the EBGP loop prevention mechanism,
the peering switches in AS11 apply an \texttt{as-path rewrite} import policy to routes from AS12 / AS13: any AS-path containing ASNs from the originating data center is replaced with ASN 11 repeated several times.

\para{The incident.} 
The operators found that the routes originated from ASN 12 could not be received by the devices in ASN 13.
The root cause is that the operators overlooked a vendor-specific behavior (VSB) of the RRs: besides the standard EBGP loop prevention, the RRs also perform IBGP loop prevention, discarding any route whose AS-path contains their own ASN.
Since $S_3$ and $S_4$ rewrote the AS-path of the AS12 route into several 11s, $RR_1$ (in AS11) treated it as an IBGP loop.
As a result, $RR_1$ could not learn the routes from AS12, and all of $RR_2$, $S_7$, $S_8$, $S_5$ and $S_6$ could not receive it.

\para{The repair given by \tool.}
\tool~gave 2026 candidate updates for the network fault. 
Among these updates, the 23$^{\textrm{rd}}$ most suspicious update is to \textit{Modify} \texttt{set as-path \textbf{replace}} to \texttt{set as-path \textbf{exclude}} in the import policy on $S_3$ for routes from $S_1$ (in AS12).
Effectively, instead of overriding the AS-path by its own ASN, it empties the AS-path attribute directly.
Thus, it stops $RR_1$'s IBGP loop prevention from triggering unintentionally, while still preventing route drops by EBGP loop prevention in other data centers.

\tool~adopted this fix, which restores reachability from AS13 to AS12 via $S_3$.
Nonetheless, AS12 is still unreachable via $S_4$, since $S_4$'s route was still being rejected.
\tool~continued searching, and found the same fix for $S_4$ in another 23 iterations.
The network is now fixed with correct ECMP paths to AS12.
In total, the repair solution is found in 118.35 seconds from 47 iterations (including 1 verification iteration at the very beginning).

\para{Lessons learned.} (1) \tool~is scalable. It can be applied on real, hyper-scale data center networks, providing repairs in minutes.
(2) SBFL in network configuration is effective, as the root cause is found in top 1.1\% of the candidate updates.

\subsection{Incident B: Route Flips}\label{subsec:incident-b}

Compared to Incident A, Incident B features a more complex and obscure phenomenon and root cause.

\para{Network and configurations.} The network is larger than Incident A's, with 10$\times$ more devices and 4$\times$ more total lines of configuration.
As shown in Figure \ref{fig:case-study}(b), the devices of AS21 originate a BGP aggregate \texttt{20.0.0.0/24} and advertise it into AS22, using an export policy that sets a community attribute.
In AS22, the switches assign a lower local preference to BGP routes coming from AS21.
Switches in AS23 re-advertise routes back into AS22 by matching community $\kappa$, which is normally set only by specific downstream devices.

\para{The incident.} After adding two new switches $S_1$ and $S_2$ in AS21, 
a route flapping happened between $S_3$, $S_4$ and $S_5$.
The root cause was: the configuration of export policy in $S_1$ and $S_2$ mistakenly set the community to $\kappa$.
Consider the propagation of \texttt{20.0.0.0/24} from $S_1$. The route propagates normally along $S_1 \rightarrow S_3 \rightarrow S_5$;
but since $S_5$ (in AS23) matches $\kappa$ in its export policy to AS22, it advertises the route back to $S_4$.
$S_4$ also receives \texttt{20.0.0.0/24} directly from $S_2$, but because that route carries the low local preference set by the import policy from AS21,
$S_4$ instead prefers the looped route from $S_5$, and withdraws the route it had advertised to $S_5$.
The same problem happens symmetrically on $S_3$.
But once $S_4$'s advertisement is withdrawn, $S_3$ again prefers the route from $S_1$.

\para{The repair given by \tool.} In this incident, \tool~does not fix the network by directly addressing the root cause.
Instead, it applies repairs at two locations, $S_3$ and $S_4$:
First, in the 5$^\textrm{th}$ iteration, \tool~finds the candidate update on $S_4$ to \textit{Modify} the import policy from AS23 from \texttt{allow} to \texttt{deny}.
Since $S_4$ can no longer receive the route from $S_5$, it prefers the route from $S_2$ despite its low local-preference.
The route from $S_4$ still matches community $\kappa$ at $S_5$ and is still preferred at $S_3$; but since it is no longer withdrawn from $S_4$, the flapping stops.

The next step is to fix the incorrect route propagation from $S_5$ to $S_3$.
In the 15$^\textrm{th}$ iteration, \tool~finds the candidate update on $S_3$ to \textit{Modify} the import policy from AS21 to a policy that sets the \textit{local-preference} of routes to 200.
This policy is grafted from the import policy of a network controller -- an SDN device in the network that also carries such policies.
As the direct cause of the incident is that $S_3$ and $S_4$'s local preference of routes received from AS21 is lower,
the repair candidate removes the condition for $S_3$ to prefer routes of \texttt{20.0.0.0/24} from $S_5$.
Finally, the route from both $S_1$ and $S_2$ is propagated correctly.
In total, the repair solution is found in 315.93 seconds from 15 iterations.

\para{Lessons learned.}
    The repairs generated by \tool~may not \textit{be} the root cause, but they can be helpful to \textit{find} the root cause.
    In this case, the lower local-preference of $S_3$ and $S_4$ is a direct cause of the route flapping problem.
    With the repair, network engineers can easily find the reason why routes in $S_5$ are sent back to $S_4$, further find the matching condition and the mistakenly set community value.
    On the other hand, one might not take the suggestion as a permanent solution for a network fault, as it might be not robust and only works for the specification when fixing.
    

\section{Related Work}
\revision{

\para{Configuration repair and synthesis.}
The closest prior work targets automatic configuration repair.
CPR~\cite{gember2017automatically} and AED~\cite{abhashkumar2020aed} synthesize repairs by encoding control-plane semantics and operator intent as SMT constraints and solving for a satisfying configuration, while CEL~\cite{gember2022localizing} localizes the root cause by solving for a minimum correction set.
Provenance-based methods~\cite{wu2014diagnosing,wu2017automated,chen2016good} trace abnormal forwarding events back to candidate root causes through a provenance graph.
Data-driven configuration management has also been explored: Auric~\cite{mahimkar2021auric} uses recommendation techniques to automatically generate cellular network configurations.
As discussed in \S\ref{subsec:semantics-driven}, all repair approaches build a formal or learned model of the network and reason over it; in contrast, \tool~adopts a \textit{syntax-driven} approach that grafts existing configurations and validates them, making it more scalable and general.

\para{Network verification, simulation, and misconfiguration detection.}
A large body of work checks configurations against intent but stops short of repairing them.
Simulators such as Batfish~\cite{fogel2015general} and verifiers such as Minesweeper~\cite{beckett2017general}, APKeep~\cite{zhang2020apkeep}, DNA~\cite{zhang2022differential}, Tiramisu~\cite{abhashkumar2019tiramisu}, VeriFlow~\cite{khurshid2012veriflow}, and Plankton~\cite{prabhu2020plankton}, among others~\cite{gember2016fast,fayaz2016efficient,ye2020accuracy,weitz2016scalable,plotkin2016scaling,zhang2022symbolic}, detect violations of forwarding properties.
Complementary data-driven techniques detect likely misconfigurations without a formal specification: SelfStarter~\cite{kakarla2020finding} infers configuration templates and flags outliers, Campion~\cite{tang2021campion} debugs configuration differences across similar routers, and pattern-mining~\cite{chauhan2021detecting} surfaces statements that deviate from common patterns.
These tools \textit{detect} or \textit{localize} errors but leave the fix to operators, which is precisely the manual burden that motivates \tool.

\para{Automatic program repair.}
\tool~adapts ideas from APR, which falls into two paradigms: semantics-driven repair that solves constraints for a correct-by-construction patch~\cite{nguyen2013semfix,mechtaev2015directfix,mechtaev2016angelix}, and syntax-driven generate-and-validate repair~\cite{le2011genprog,liu2013r2fix,wei2010automated,long2016automatic} built on the plastic surgery hypothesis~\cite{barr2014plastic,xia2023plastic}; we refer the reader to surveys~\cite{gazzola2018automatic,monperrus2018automatic} for a broader treatment.
Our localization step builds on spectrum-based fault localization (SBFL)~\cite{jones2005empirical,agarwal2014fault,abreu2007accuracy,wong2012software}, which scores a statement's suspiciousness from its coverage by passing and failing tests.
To our knowledge, \tool~is the first to adapt syntax-driven APR and SBFL to network configuration repair.

}

\section{Discussion}

\para{Using LLMs for fix generation.}
In this paper, we leveraged the ``plastic surgery hypothesis'' for repairing network configurations.
When generating fixes, we used heuristics to generate fixes from candidates.
Recent studies (e.g., \cite{xia2023plastic}) suggests that LLMs can be used to find better fixes. 
We will try to apply similar approaches to improve the quality of fixes in the future.

\para{Repairing data plane misconfigurations.}
\tool{} currently repairs only misconfigurations for routing protocols (e.g., BGP).
There are other configurations used for data plane forwarding (e.g., policy-based routing and NATs), 
and can also cause network incidents if there are errors. 
For data-plane configuration, the same SBFL approach also works.
By viewing packets as test cases, and test packet forwarding properties (reachability, blackhole-freedom, etc.),
the localization approach can be also applied on data plane configurations.
As the simulator only supports simulation of the control plane, \tool~currently is unable to fix these configurations.
In the future, we will try to implement the ACR principle on data plane or general network simulators to enable the repair of data plane configurations.


\para{Repairing wide-area network configurations.}
In this paper, we design and evaluate \tool~based on data center network configurations.
For other kinds of networks such as WANs, we tentatively think the plastic surgery hypothesis also holds, as there are also some level of redundancy (e.g. multiple border routers peering with the same ASes).
Therefore, we hypothesize \tool~can be a general tool for repairing these networks.
We leave the evaluation of \tool~on WANs as a future work.



\para{Limitations.}
For those incidents caused by missing the origin of the route -- advertisement statement, route redistribution, static route, etc.,
\tool~cannot generate the coverage report based on current test suite, since the faulty prefixes are not generated, and configuration units cannot.
We leave it to manual troubleshooting because they are relatively obvious.




\section{Conclusion}

In this paper, we motivate the need of automatic configuration repair (ACR) with 
experiences of real-world network incidents.
Inspired by APR, we propose to achieve ACR with a syntax-driven approach, which is general and scalable.
Specifically, we design and implement \tool, and evaluate it based on configurations of real production networks.
We show \tool{} can repair over 97.5\% of the injected errors within an average of 6.93 seconds.

\noindent\textit{This work does not raise any ethical issues.}

\bibliographystyle{plain}
\bibliography{main}

\end{document}